\documentclass[twocolumn,aps,prb,showpacs,superscriptaddress,longbibliography,10pt,floatfix]{revtex4-2} 
\usepackage{amsfonts}
\usepackage{amsmath}
\usepackage{amssymb}
\usepackage{graphicx}
\usepackage{placeins}
\usepackage[colorlinks=true,linkcolor=blue,citecolor=blue]{hyperref}

\begin{document}
\title{An approximate analytical method for quasiperiodic chains with integer-power hopping modulation}


\author{Ni Lu}
\affiliation{Department of Physics, Jiangsu University, Zhenjiang, 212013, China}
\author{Xianqi Tong} \altaffiliation{xqtong@ujs.edu.cn}
\affiliation{Department of Physics, Jiangsu University, Zhenjiang, 212013, China}

\date{\today}

\begin{abstract}
Quasiperiodic modulation of nearest-neighbor hopping can produce different localization responses across the spectrum even without an onsite potential. We study chains whose hopping amplitudes are integer powers of a shifted cosine and develop an analytical framework based on two exact constructions. On an odd open chain, the zero-mode recurrence yields bounded logarithmic amplitude fluctuations when the hopping function is bounded away from zero; the same recurrence defines a rescaling in which the finite-energy spatial modulation is proportional to energy. When the hopping function develops zeros, the zero-mode fluctuations become unbounded for chains with nonzero sampled bonds. Independently, eliminating one sublattice gives an exact Jacobi equation. A first-harmonic truncation of this equation provides a characteristic scale for changes in the outer-spectrum states and predicts an inverse-power dependence at large hopping exponents. Finite-size participation ratios and transfer-matrix Lyapunov exponents support distinct responses of central and outer states, while also delimiting the range of the approximations.
\end{abstract}

\maketitle

\section{Introduction}

Quasiperiodic lattices provide a setting in which deterministic spatial modulation can produce extended, localized, and critical single-particle states \cite{2,3,4,5,6,7,8,10,50,54,55,89}. Topological band theory in periodic systems provides part of the conceptual foundation for related studies in aperiodic settings \cite{1}. Localization, transport, and topology have been investigated in optical-lattice studies and diffraction experiments \cite{10,37,40,88,90}, photonic structures \cite{9,39,41,42,43,87,97}, electronic quasicrystals and their transport properties \cite{38,45,46,47,48,50,62}, mechanical realizations \cite{93}, and systems involving magnetic fields or magnonic quasiperiodicity \cite{44,49,53}. General non-Hermitian symmetry, topology, and skin effects provide a broader framework \cite{63,64,65,66,67,68}; dephasing and non-Hermitian quasiperiodic models then show how localization can be modified by dissipation, gain and loss, or nonreciprocal motion \cite{51,52,71,72,73,74,75,76,86,98,99,100,101,102}. In the Aubry--Andr\'e--Harper (AAH) model, self-duality gives an energy-independent localization transition \cite{2,3,105}. Generalized quasiperiodic potentials or couplings can break this structure and produce mobility edges, so that states at different energies have different localization properties at the same modulation strength \cite{11,12,13,14,15,16,17,19,20,21,22,23,92,94,95,96}. Mobility edges can also occur through distinct mechanisms in engineered multiband or flat-band lattices \cite{35}. Analytical localization boundaries are nevertheless available only for selected models. They commonly rely on a special duality, an exactly tractable Lyapunov exponent, or another model-specific construction \cite{14,17,19,20,21,91,106}. Approximate approaches are therefore useful for more complicated quasiperiodic profiles, provided that their assumptions and range of validity are distinguished from exact results \cite{18}.

Most mobility-edge mechanisms rely on an onsite quasiperiodic potential, whereas bond modulation acts directly on the kinetic couplings. In off-diagonal AAH, dimerized, and mosaic chains, varying the bond pattern alters localization transitions, dynamics, and spectral properties \cite{28,29,30,31,32,33,34}, alongside topological phases protected by chiral or crystalline symmetries \cite{32,33,36} and related \(p\)-wave pairing extensions \cite{31,103}. Without an onsite term, such bipartite chains also obey chiral symmetry, which pairs nonzero energies at \(\pm E\) and gives the zero-energy sector a distinct real-space structure \cite{56,57,58,59,60,61}. The model studied here assigns to every nearest-neighbor bond an integer power of a shifted cosine: unlike Hermitian and non-Hermitian models with distance-dependent power-law hopping \cite{24,25,26,27,69,70}, the coupling remains strictly short ranged, while increasing the exponent reshapes the local profile, introduces additional Fourier harmonics, and enhances bond contrast. Understanding how this single nearest-neighbor modulation acts differently on the central and outer parts of the spectrum is the focus of this work.

Here a chain with integer-power hopping modulation is used to develop a practical analytical framework for this problem. The purpose is not to assign an exact transition boundary from a single approximation, but to combine exact transformations with controlled interpretations of the resulting estimates. The zero-mode recurrence on an odd open chain gives an exact product solution and, when the hopping function is bounded away from zero, an exact rescaling of the finite-energy equation. In the transformed equation, the spatially varying term is proportional to energy, which provides a direct mechanism for the weaker response of states near zero energy. Phase averaging then gives a local estimate of the zero-energy density of states. Independently, eliminating one sublattice produces an exact Jacobi equation whose hopping and diagonal coefficients inherit correlated quasiperiodic modulations from the original bonds. A first-harmonic truncation of this equation defines a characteristic modulation scale for the outer states, while a model density of states gives an estimate of the spectral edge. These two steps are approximate and are not identified with exact mobility edges of the original chain. Their implications and limitations are assessed using finite-size participation ratios and transfer-matrix Lyapunov exponents. Because the construction is expressed in terms of the hopping function and its Fourier components, the same procedure can also be examined for other quasiperiodic hopping profiles.

Section~\ref{sec:model} introduces the model, its chiral symmetry, and the localization diagnostics. Section~\ref{sec:p2_overview} presents the finite-size spectra for quadratic modulation and derives the exact zero mode of an odd open chain. Section~\ref{sec:Eneq0} develops the rescaled low-energy equation and the Jacobi-based spectral estimates. Section~\ref{sec:powerlaw} examines how the first-harmonic scale varies with the hopping exponent. Section~\ref{sec:conclusions} summarizes the results and their limitations.

\begin{figure}[t]
\centering
\includegraphics[width=1\linewidth]{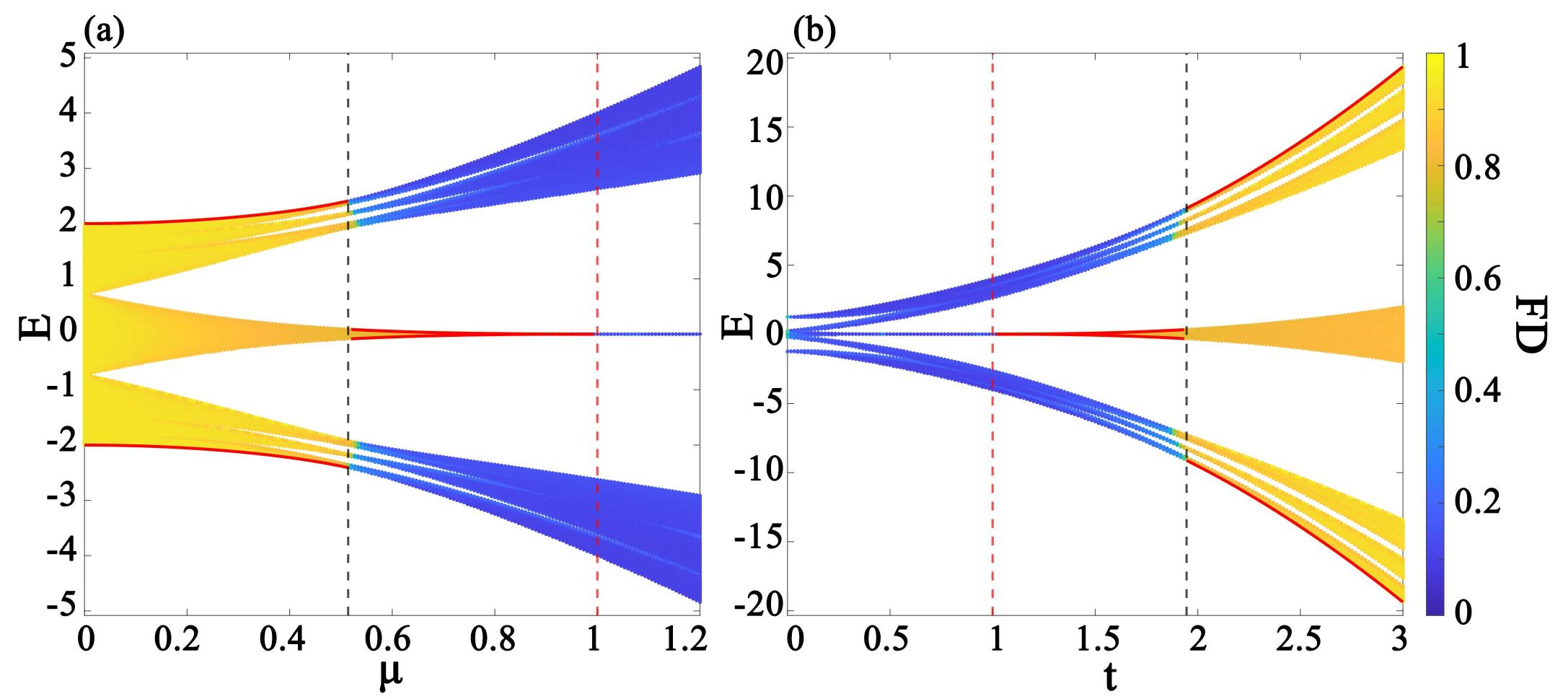}
\caption{
Finite-size spectra for quadratic hopping modulation on an even periodic Fibonacci ring with \(L=610\), \(\alpha=377/610\), and \(\phi=0\). The color denotes the finite-size fractal dimension. (a) Varying \(\mu\) at fixed \(t=1\). (b) Varying \(t\) at fixed \(\mu=1\). Black dashed lines mark the first-harmonic scale \(\mu_c\) and its equivalent value on the \(t\) cut. Red dashed lines mark the hopping-zero condition \(\mu=t\). Where shown, red solid curves denote the approximate outer spectral edge \(\pm E_{\mathrm{sp}}\) and the central energy scale \(\pm W_c\).
}
\label{fig:ME_new}
\end{figure}
\section{Model and symmetry}
\label{sec:model}

Consider a Hermitian tight-binding chain with real nearest-neighbor hopping and no onsite potential. The single-particle Hamiltonian on an open chain of \(L\) sites is
\begin{equation}
H=\sum_{n=1}^{L-1}J_n
\left(c_n^\dagger c_{n+1}+c_{n+1}^\dagger c_n\right),
\label{eq:H}
\end{equation}
with hopping amplitudes
\begin{equation}
J_n=j(\theta_n)^p,\qquad
j(\theta)=t+\mu\cos\theta,\qquad
\theta_n=2\pi\alpha n+\phi .
\label{eq:hopping}
\end{equation}
Here \(t>0\), \(\mu\geq0\), and \(p\) is a positive integer. The irrational frequency is the golden mean, \(\alpha=(\sqrt5-1)/2\). The continuous function \(J(\theta)=j(\theta)^p\) defines the hopping profile, and the sequence of phases \(\theta_n\) densely samples the phase circle. The spectra shown below use \(p=2,3,\) and \(4\); the analytical modulation scale is evaluated for \(p=2,\ldots,8\).

On an open chain each bond connects an odd site to an even site. The sublattice operator \(\Gamma|n\rangle=(-1)^n|n\rangle\) therefore gives
\begin{equation}
\Gamma H\Gamma^{-1}=-H.
\label{eq:chiral}
\end{equation}
The spectrum of this bipartite chain is symmetric about zero energy \cite{32,33,56,57}. An odd open chain has one more odd site than even sites, which guarantees a zero mode. Under periodic boundary conditions, the closing bond preserves the bipartite structure only for even \(L\). The spectral figures and finite-size checks below therefore use even Fibonacci rings, whereas the exact zero-mode analysis uses odd open chains.

For \(p=2\), the hopping contains a constant part and two Fourier harmonics:
\begin{equation}
J_n=t^2+\frac{\mu^2}{2}
+2t\mu\cos\theta_n+\frac{\mu^2}{2}\cos(2\theta_n).
\label{eq:squared_expansion}
\end{equation}
Varying \(\mu\) then changes both the mean hopping and the strength of its modulation.

For a normalized eigenstate \(\psi^{(\nu)}\), its inverse participation ratio (IPR) and finite-size fractal dimension (FD) are defined as
\begin{equation}
\mathrm{IPR}=\sum_{n=1}^{L}|\psi_n^{(\nu)}|^4,\qquad
\mathrm{FD}=-\frac{\ln\mathrm{IPR}}{\ln L}.
\label{eq:IPR_FD}
\end{equation}
At fixed \(L\), larger FD means that the state occupies more sites in the sense of the IPR. The size dependence distinguishes the limiting cases. Extended states have an IPR of order \(L^{-1}\), and exponentially localized states with a finite localization length have an IPR that approaches a nonzero value as \(L\) grows. We use \(E\) for the energy of the original chain and \(\lambda=E^2\) for the spectral variable of the reduced Jacobi equation.

\section{Finite-size spectra and the exact zero mode}

\label{sec:p2_overview}

Fig.~\ref{fig:ME_new} shows the eigenvalues and FD of an even periodic Fibonacci approximant with \(L=610\), \(\alpha=377/610\), and \(\phi=0\). Panel (a) varies \(\mu\) at fixed \(t=1\), and panel (b) varies \(t\) at fixed \(\mu=1\). Writing \(x=\mu/t\) gives \(J_n=t^p[1+x\cos\theta_n]^p\). At fixed \(p\), the eigenvectors and FD depend on \(x\), while the energies rescale by \(t^p\). The two panels therefore display the same parameter dependence on different energy scales.

As \(x\) increases, the FD decreases first near the outer spectral edges, while states with relatively large FD remain in a narrower interval around zero. The size dependence is checked by ordering the positive energies of even Fibonacci rings and comparing fixed-rank windows in the central and outer parts of the spectrum. Between \(L=610\) and \(2584\), the outer-state IPR follows extended scaling at \(x=0.50\) but becomes nearly size independent at \(x=0.52\). Over the same sizes, the selected central states retain approximately inverse-size IPR scaling.

The transfer-matrix calculation gives the same distinction. At \(x=0.52\), the finite-length Lyapunov exponent approaches a positive value in the edge window, whereas the central-window value continues to decrease with propagation length. Together, the two diagnostics support localized outer states and extended scaling in the selected central window at the same modulation strength. Appendix~\ref{sec:supp_comparison} gives the numerical protocol and convergence checks.

The levels near zero in the periodic spectra motivate a closer look at the state at exactly zero energy. For this calculation we use an odd open chain (\(L=2M+1\)), where sublattice imbalance guarantees a zero mode. With nonzero sampled hoppings, this mode is unique and resides on the odd sublattice. The zero-energy equation at site \(2m\) gives \(\psi_{2m+1}=-(J_{2m-1}/J_{2m})\psi_{2m-1}\). Iteration gives the logarithmic amplitude relative to the first site,
\begin{equation}
S_m\equiv \ln\left|\frac{\psi_{2m+1}}{\psi_1}\right|
=\sum_{k=1}^{m}\ln\left|\frac{J_{2k-1}}{J_{2k}}\right|.
\label{eq:log_zero_mode}
\end{equation}
We measure its spatial variation by the range \(R(L)=\max_m S_m-\min_m S_m\).

For \(\mu<t\), the hopping function is bounded away from zero. At the golden-mean frequency, the sum in Eq.~\eqref{eq:log_zero_mode} reduces to the difference of two values of a bounded function (Appendix~\ref{sec:supp_gauge}). Thus \(R(L)\) remains bounded and the inverse participation ratio of the normalized zero mode scales as \(L^{-1}\). The hopping function first vanishes at \(\mu=t\). At and above this value, the logarithm of an adjacent hopping ratio can become arbitrarily large in magnitude as the chain samples phases near a zero. Consequently, \(R(L)\) grows without bound as \(L\) increases (Appendix~\ref{sec:supp_gauge}). At fixed \(\mu/t\), \(\alpha\), and \(\phi\), \(S_m\) is proportional to \(p\).

\begin{figure}[tbp]
\centering
\includegraphics[width=1\linewidth]{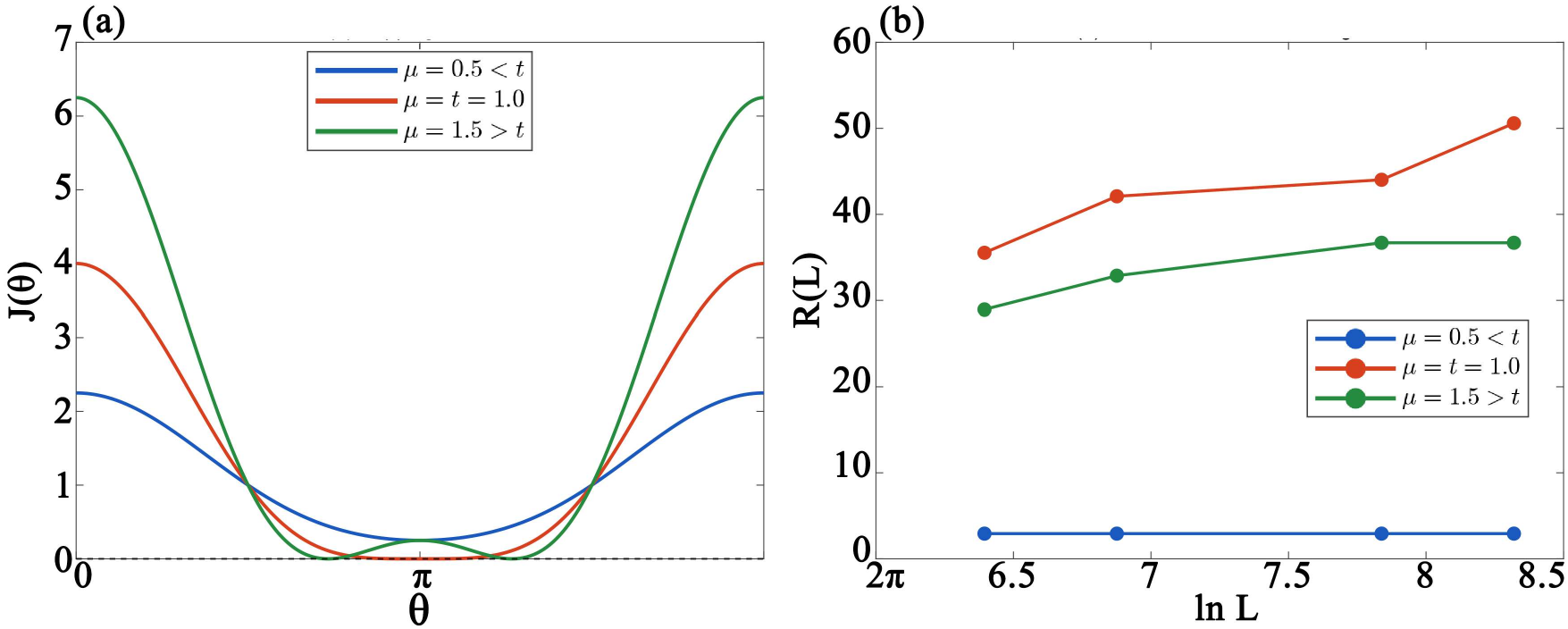}
\caption{
Zero-energy state of the quadratically modulated chain at \(t=1\), \(\alpha=(\sqrt{5}-1)/2\), and \(\phi=0\). (a) Hopping function \(J(\theta)=[t+\mu\cos\theta]^2\) for \(\mu=0.5\), \(1.0\), and \(1.5\). (b) Range \(R(L)\) of the logarithmic zero-mode amplitude versus \(\ln L\) for odd-length open chains at the same modulation strengths.
}
\label{fig:zero_mode_scaling}
\end{figure}

Fig.~\ref{fig:zero_mode_scaling} illustrates the two regimes for \(p=2\). Over the sizes shown, \(R(L)\) is nearly constant below the hopping-zero threshold and increases once zeros appear. The bounded zero-mode solution for \(\mu<t\) also transforms the equation at nonzero energy and identifies how energy enters the spatial modulation.

\section{Approximate analytical method}

\label{sec:Eneq0}

The zero-mode transformation makes the energy dependence of the spatial modulation explicit and gives a way to estimate the density of states at zero energy. We then derive a Jacobi equation for the nonzero-energy states and use it to estimate the modulation scale seen at the spectral edge.

\subsection{Energy dependence and zero-energy density of states}

For \(0\leq\mu<t\), let \(g_n=e^{\chi_p(\theta_{n-1})}\), where \(\chi_p\) is the bounded, zero-mean function constructed in Appendix~\ref{sec:supp_gauge}. With \(q_0=\langle\ln J\rangle\), it satisfies \(J_ng_ng_{n+1}=e^{q_0}\). Here \(\langle\cdot\rangle\) denotes the phase average. Substituting \(\psi_n=g_n\eta_n\) into the eigenvalue equation gives
\begin{equation}
\eta_{n+1}+\eta_{n-1}=E b_n\eta_n,\qquad b_n=e^{-q_0}g_n^2.
\label{eq:low_energy_gauge}
\end{equation}
At fixed \(\mu<t\), \(g_n\) is bounded above and away from zero. The rescaling therefore preserves exponential decay and changes the IPR of a normalized state by at most size-independent factors. Writing \(C_p=\langle b\rangle=e^{-q_0}\langle e^{2\chi_p}\rangle\), we can separate \(Eb_n\) into a constant term \(EC_p\) and a spatially varying term \(E(b_n-C_p)\). The latter weakens as \(E\) approaches zero at fixed hopping profile. This energy dependence explains why the central states can respond more weakly to the same bond modulation.

Replacing \(b_n\) by its phase average \(C_p\) gives a uniform chain with hopping \(1/C_p\) and a zero-energy density of states per site of \(C_p/(2\pi)\). This replacement is a local low-energy approximation, not a controlled description of the full central spectrum. To infer a corresponding energy scale, the central group of levels is modeled by an arcsine density on \([-W_c,W_c]\) with total weight \(\delta_0=\|2\alpha\|\), where \(\|x\|\) is the distance to the nearest integer. This weight is an allowed gap label for the even-sublattice frequency \(2\alpha\) \cite{DamanikFillmanZhang2023} and agrees with the central spectral weights found in Appendix~\ref{sec:supp_comparison}. Matching the model density at zero energy to the phase-average estimate gives
\begin{equation}
W_c=\frac{2\delta_0}{C_p}.
\label{eq:Wc}
\end{equation}
As \(\mu\to t^-\), the average \(\langle J^{-1}\rangle\) diverges. The bound \(C_p\geq\langle J^{-1}\rangle\) in Appendix~\ref{sec:supp_gauge} therefore implies that this scale tends to zero.

For quadratic modulation, the resulting scale agrees with the central half-width at moderate modulation but substantially underestimates it close to the hopping-zero threshold. A direct finite-window count nevertheless remains consistent with the local zero-energy density estimate in that regime (Appendix~\ref{sec:supp_comparison}). Thus the phase average is useful for the local density of states, whereas extending it across the entire central group through the arcsine model has a more limited range of validity.

\subsection{Jacobi reduction and spectral estimates}

\label{sec:mapping}
\label{sec:mu_FH}

Consider an open chain with \(L=2M\) sites and denote the odd- and even-site amplitudes by \(u_m=\psi_{2m-1}\) and \(v_m=\psi_{2m}\). For \(E\ne0\), the equation at site \(2m-1\) gives \(u_m=(J_{2m-2}v_{m-1}+J_{2m-1}v_m)/E\). Substituting this expression into the equation at site \(2m\) gives
\begin{equation}
a_{m-1}v_{m-1}+V_mv_m+a_mv_{m+1}=\lambda v_m,
\label{eq:Jacobi}
\end{equation}
where \(\lambda=E^2\) and
\begin{equation}
a_m=J_{2m}J_{2m+1},\qquad V_m=J_{2m-1}^2+J_{2m}^2.
\label{eq:aV_def}
\end{equation}
Terms involving bonds outside the open chain are omitted. For an even periodic chain, the same reduction holds with cyclic indexing. Appendix~\ref{sec:supp_esp} gives the matrix form of the reduction.

The effective hopping \(a_m\) comes from two successive hops from site \(2m\) to site \(2m+2\). The diagonal term \(V_m\) comes from hops to either neighbour and back. Both coefficients are therefore determined by the same quasiperiodic sequence of bonds. The even sites sample the modulation at \(\theta_{2m}=2m\omega+\phi\), where \(\omega=2\pi\alpha\). We write \(a_m=a(\theta_{2m})\) and \(V_m=V(\theta_{2m})\), with \(a(\theta)=j(\theta)^p j(\theta+\omega)^p\) and \(V(\theta)=j(\theta)^{2p}+j(\theta-\omega)^{2p}\). Both functions can contain Fourier harmonics up to order \(2p\).

To estimate the modulation scale, we keep the mean hopping \(a_0=\langle a\rangle\) and approximate the diagonal term by its mean \(V_0=\langle V\rangle\) and its first harmonic. With \(\widetilde V_1=\langle V(\theta)e^{-i\theta}\rangle\), the truncated Jacobi equation becomes
\begin{equation}
\begin{aligned}
&a_0(v_{m+1}+v_{m-1})
+2|\widetilde V_1|\cos(2m\omega+\phi')v_m
\\
&\qquad=(\lambda-V_0)v_m,
\end{aligned}
\label{eq:AAH_form}
\end{equation}
where \(\phi'=\phi+\arg\widetilde V_1\). This is the Aubry--Andr\'e--Harper model, with \(V_0\) setting the origin of the spectral variable \cite{105,2,3}. Its self-dual point defines the first-harmonic scale \(\mu_c\) through
\begin{equation}
|\widetilde V_1(p,\mu_c)|=a_0(p,\mu_c).
\label{eq:mu_FH_condition}
\end{equation}

This condition defines a characteristic scale of the truncated equation, not an exact localization boundary of the original Jacobi operator. It neglects both the spatial variation of \(a(\theta)\) and the higher harmonics of \(V(\theta)\), whose magnitudes are not controlled by a small parameter at the resulting scale. For quadratic modulation, \(\mu_c\) lies between the two sampled strengths in Sec.~\ref{sec:p2_overview}, where the outer states show different size dependence. The estimate therefore identifies the relevant modulation range, while the original-chain diagnostics establish the energy dependence.

We also use the Jacobi equation to estimate the outer spectral edge \(E_{\mathrm{sp}}\). In \(\lambda=E^2\), the model density places a weight \(\delta_0\) at zero and spreads the remaining weight over one outer interval. Its endpoints are fixed by matching the first moment and setting the logarithmic potential to the value for a vanishing Lyapunov exponent in the Thouless formula~\cite{104}. Appendix~\ref{sec:supp_esp} gives the resulting expression for the outer spectral edge. This estimate concerns the extent of the spectrum in energy.

\section{Dependence of the modulation scale on the hopping exponent}

\label{sec:powerlaw}

\begin{figure}[t]
\centering
\includegraphics[width=1.0\linewidth]{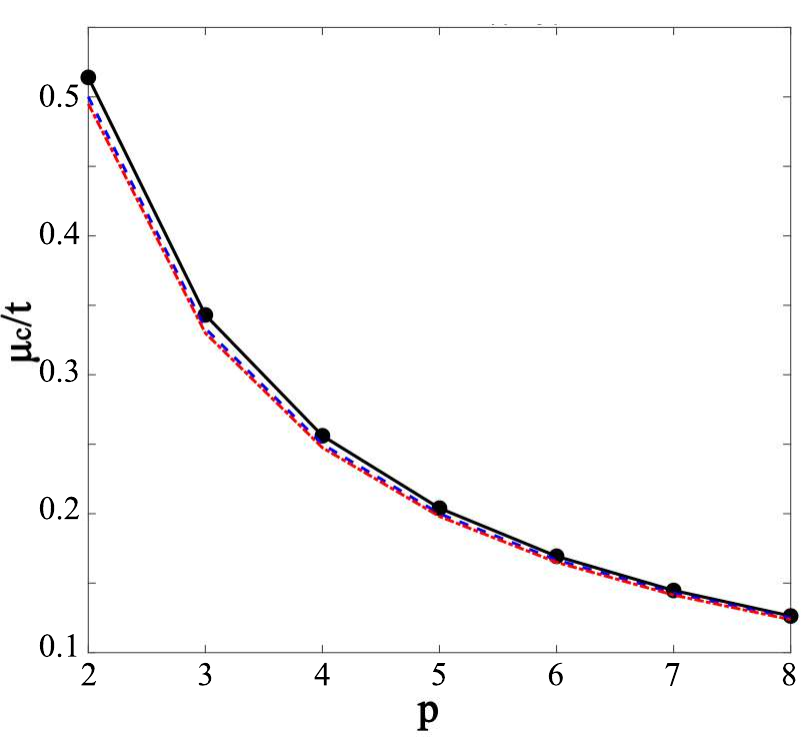}
\caption{
First-harmonic scales \(\mu_c/t\) for \(p=2,\ldots,8\) (black circles). The blue dashed line is a \(1/p\) guide, and the red dash-dotted line is the large-\(p\) asymptote \(C_\infty/p\), with \(C_\infty\simeq0.9908\).
}
\label{fig:powerlaw_mu_c}
\end{figure}

Fig.~\ref{fig:powerlaw_mu_c} shows that the first-harmonic scale \(\mu_c/t\) decreases as the hopping exponent \(p\) increases. Both coefficients in the balance condition of Sec.~\ref{sec:mu_FH} contain a factor \(t^{2p}\). This factor cancels, so the estimated ratio depends only on \(p\) at fixed quasiperiodic frequency. Appendix~\ref{sec:supp_fourier} gives the Fourier coefficients used for \(p=2,\ldots,8\).

\begin{figure*}[t]
\centering
\includegraphics[width=1.0\linewidth]{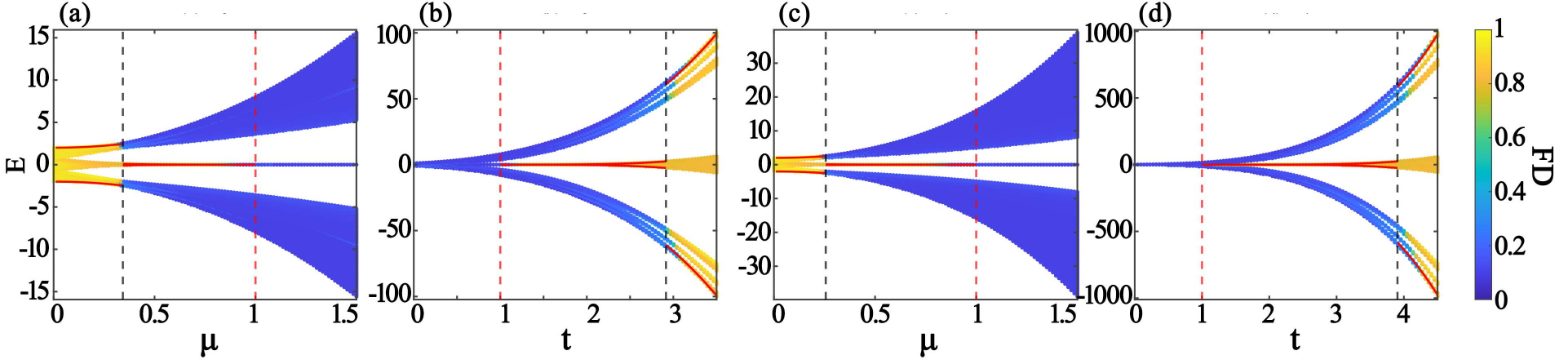}
\caption{
Finite-size spectra for \(p=3\) [(a), (b)] and \(p=4\) [(c), (d)] on an even periodic Fibonacci ring with \(L=610\), \(\alpha=377/610\), and \(\phi=0\). Panels (a) and (c) vary \(\mu\) at fixed \(t=1\), and panels (b) and (d) vary \(t\) at fixed \(\mu=1\). The color denotes FD. Black dashed lines mark the first-harmonic scales, and red dashed lines mark \(\mu=t\). Where shown, red solid curves denote the approximate outer spectral edge and the central energy scale inferred from the zero-energy density of states.
}
\label{fig:phase_p34}
\end{figure*}

The \(1/p\) dependence follows by setting \(\mu/t=C/p\). The normalized hopping \(J(\theta)/t^p=[1+(C/p)\cos\theta]^p\) then approaches \(e^{C\cos\theta}\) as \(p\to\infty\). Substituting this limiting form into the self-duality condition gives (Appendix~\ref{sec:supp_fourier})
\begin{equation}
2c_\alpha I_1(2C)=I_0(2c_\alpha C),
\label{eq:large_p_closure}
\end{equation}
where \(c_\alpha=|\cos(\pi\alpha)|\) and \(I_n\) is the modified Bessel function of the first kind. For the golden-mean frequency, the limiting equation has a simple positive root \(C_\infty\simeq0.9908\). The first-harmonic estimate therefore satisfies
\begin{equation}
\frac{\mu_c}{t}=\frac{C_\infty}{p}+O(p^{-2}).
\label{eq:mu_FH_asymptotic}
\end{equation}

Figure~\ref{fig:phase_p34} compares the first-harmonic scales with finite-size spectra for \(p=3\) and \(4\). Both the analytical scale and the change in the outer-state FD move to weaker modulation as \(p\) increases. Their detailed locations do not coincide, as expected because the full chain retains the spatial modulation of \(a(\theta)\) and all harmonics of \(V(\theta)\). The figure should therefore be read as a test of the trend and parameter range, not as a numerical verification of an exact transition point.

\FloatBarrier
\section{Conclusions}

\label{sec:conclusions}

For quadratic hopping modulation, participation ratios and transfer-matrix Lyapunov exponents distinguish the responses of the central and outer parts of the spectrum at the same modulation strength. The exact zero-mode transformation explains this energy selectivity: in the nonsingular regime, the spatially varying term in the transformed equation is proportional to energy and therefore weakens near zero. On an odd open chain, the same construction gives bounded logarithmic zero-mode fluctuations while the hopping function remains nonzero; once zeros appear, those fluctuations become unbounded for chains with nonzero sampled bonds.

The exact sublattice reduction also provides a systematic starting point for spectral estimates. Its first-harmonic truncation identifies a characteristic modulation scale for the outer states and yields an inverse-power trend at large hopping exponents. Because the truncation discards sizable hopping variations and higher harmonics, this scale is not an exact localization boundary. Likewise, the phase-averaged zero-energy density is more reliable as a local quantity than as an estimate of the entire central bandwidth. These limitations separate the exact structural results from the model-dependent approximations and indicate how the framework can be tested for other quasiperiodic hopping profiles.

\begin{acknowledgments}
This work was supported by the National Natural Science Foundation of China (Grant No.~ZK20260179).
\end{acknowledgments}

\section*{Data availability}

The data that support the findings of this article are available from the authors upon reasonable request.

\appendix
\renewcommand{\appendixname}{APPENDIX}

\section{Sublattice reduction and spectral-edge estimate}

\label{sec:supp_esp}

In the sublattice basis \((u,v)\) of Sec.~\ref{sec:mapping}, the Hamiltonian of an even open chain takes the form
\begin{equation}
H=\begin{pmatrix}0&Q\\Q^\dagger&0\end{pmatrix},\qquad
Q_{m\ell}=J_{2m-2}\delta_{\ell,m-1}+J_{2m-1}\delta_{\ell,m}.
\label{eq:H_block}
\end{equation}
We set \(J_0=J_{2M}=0\) at the open ends. Eliminating \(u\) from \(Qv=Eu\) and \(Q^\dagger u=Ev\) gives \(Q^\dagger Qv=E^2v\) for \(E\ne0\). The action of this operator on \(v\) is
\begin{equation}
\begin{aligned}
(Q^\dagger Qv)_m={}&J_{2m-2}J_{2m-1}v_{m-1}\\
&+(J_{2m-1}^2+J_{2m}^2)v_m
+J_{2m}J_{2m+1}v_{m+1}.
\end{aligned}
\label{eq:QdagQv}
\end{equation}
Taking \(\lambda=E^2\) gives Eq.~\eqref{eq:Jacobi}.

For the spectral-edge estimate of Sec.~\ref{sec:mapping}, we approximate a narrow and well-separated central group by a point mass with weight \(\delta_0\). We model the rest of the density-of-states measure by a continuous density on a single interval,
\begin{equation}
dN_{\mathrm{app}}(\lambda)
=\delta_0\,\delta(\lambda)\,d\lambda
+\rho_{\mathrm{out}}(\lambda)\,d\lambda ,
\label{eq:point_mass}
\end{equation}
where \(\rho_{\mathrm{out}}\) is supported on \([\lambda_-,\lambda_+]\). With \(\lambda_\pm=(A\pm B)^2\) and \(A>B>0\), define
\begin{equation}
\begin{aligned}
\mathcal R(z)&=\sqrt{(z-\lambda_-)(z-\lambda_+)},\\
P&=\sqrt{\lambda_-\lambda_+}=A^2-B^2 ,
\end{aligned}
\label{eq:P_R}
\end{equation}
using the branch \(\mathcal R(z)\sim z\) at infinity. We choose a Cauchy transform with \(z\) in the denominator for the pole at zero and \(\mathcal R(z)\) for the single interval. A monic linear numerator ensures the \(1/z\) decay of a normalized measure:
\begin{equation}
G(z)=\frac{z-c}{z\mathcal R(z)} .
\label{eq:Cauchy}
\end{equation}
Since \(\mathcal R(0)=-P\), the residue at zero is \(c/P\). Setting this residue equal to \(\delta_0\) fixes \(c=\delta_0P\). The Stieltjes inversion formula then gives
\begin{equation}
\rho_{\mathrm{out}}(\lambda)
=\frac{\lambda-c}
{\pi\lambda\sqrt{(\lambda-\lambda_-)(\lambda_+-\lambda)}},
\quad \lambda_-<\lambda<\lambda_+ .
\label{eq:rho_out}
\end{equation}
The outer interval carries weight \(1-\delta_0\). Its density is nonnegative when \(c\leq\lambda_-\), or equivalently
\begin{equation}
\frac BA\leq \frac{1-\delta_0}{1+\delta_0}.
\label{eq:density_positivity}
\end{equation}
This condition selects admissible solutions for the model measure. On the interior of the outer interval the real part of \(G(\lambda+i0)\) vanishes. It is the principal-value derivative of the logarithmic potential, so the potential is constant on this interval.

The endpoints are fixed by two conditions. For the first, expanding \(G(z)\) at infinity gives the first moment as the coefficient of \(z^{-2}\),
\begin{equation}
\begin{aligned}
\int\lambda\,dN_{\mathrm{app}}(\lambda)
&=\frac{\lambda_-+\lambda_+}{2}-c\\
&=(1-\delta_0)A^2+(1+\delta_0)B^2 .
\end{aligned}
\label{eq:m1_AB}
\end{equation}
The first moment of the exact Jacobi density of states is the phase average of the diagonal term, \(\langle V\rangle=2\langle j^{2p}\rangle\). Matching the two moments gives
\begin{equation}
(1-\delta_0)A^2+(1+\delta_0)B^2=2\langle j^{2p}\rangle .
\label{eq:trace_closure}
\end{equation}

For the infinite Jacobi chain, the Thouless formula \cite{104} relates the Lyapunov exponent to the exact density-of-states measure \(dN\):
\begin{equation}
\gamma(\lambda)=
\int\ln|\lambda-\lambda'|\,dN(\lambda')-\langle\ln|a|\rangle .
\label{eq:Thouless}
\end{equation}
To determine the constant value of the potential for our model measure, we set it equal to \(\langle\ln|a|\rangle\), the value corresponding to a vanishing Lyapunov exponent. Write \(\rho_{\mathrm{out}}(\lambda)=(1-c/\lambda)\sigma(\lambda)\), where \(\sigma\) is the arcsine density on \([\lambda_-,\lambda_+]\). At the upper endpoint the potential is
\begin{equation}
U_{\mathrm{app}}(\lambda_+)
=\delta_0\ln\lambda_++\mathcal J_0-c\mathcal J_1 ,
\label{eq:Ub_J}
\end{equation}
where \(\mathcal J_0\) and \(\mathcal J_1\) are the integrals of \(\ln(\lambda_+-\lambda)\) and \(\lambda^{-1}\ln(\lambda_+-\lambda)\), respectively, against \(\sigma\). With \(r=B/A\), the substitution \(\lambda=A^2+B^2+2AB\cos\theta\) gives
\begin{equation}
\mathcal J_0=\ln(AB),\qquad
\mathcal J_1=\frac{\ln(\lambda_+r)}{A^2-B^2}.
\label{eq:J_AB_final}
\end{equation}
The first identity follows by averaging \(\ln(\lambda_+-\lambda)=\ln(AB)+\ln|1-e^{i\theta}|^2\). In the second integral the factor \(P/\lambda\) is the Poisson kernel at \(-r\), which replaces \(\ln|1-e^{i\theta}|^2\) by \(2\ln(1+r)\) in the average. Substituting \(c=\delta_0(A^2-B^2)\) into Eq.~\eqref{eq:Ub_J} yields
\begin{equation}
U_{\mathrm{app}}(\lambda_+)=\ln(AB)+\delta_0\ln(A/B).
\label{eq:Ub_final}
\end{equation}
For \(0\leq\mu<t\), the hopping function is positive and \(\langle\ln|a|\rangle=2p\ln[(t+\sqrt{t^2-\mu^2})/2]\). This phase average also follows from the factorization in Appendix~\ref{sec:supp_gauge}. Equating it to Eq.~\eqref{eq:Ub_final} gives
\begin{equation}
A^{1+\delta_0}B^{1-\delta_0}
=\left(\frac{t+\sqrt{t^2-\mu^2}}{2}\right)^{2p}.
\label{eq:thouless_closure}
\end{equation}

To solve Eqs.~\eqref{eq:trace_closure} and~\eqref{eq:thouless_closure}, let \(K=[(t+\sqrt{t^2-\mu^2})/2]^{2p}\) and use \(r=B/A\). Eliminating \(A\) gives
\begin{equation}
Kr^{\delta_0-1}\big[(1-\delta_0)+(1+\delta_0)r^2\big]
=2\langle j^{2p}\rangle .
\end{equation}
On the range \(0<r\leq(1-\delta_0)/(1+\delta_0)\) the left-hand side is decreasing. From the root we obtain \(A=\sqrt{Kr^{\delta_0-1}}\), \(B=Ar\), and the estimate \(E_{\mathrm{sp}}=A+B\).

\section{Zero-energy wave function and low-energy estimate}

\label{sec:supp_gauge}

For an odd open chain with nonzero sampled hoppings, the left boundary equation at \(E=0\) gives \(\psi_2=0\). The equations at successive odd sites then give \(\psi_{2m}=0\) throughout the chain. The equations at the even sites determine every odd-site amplitude in terms of \(\psi_1\):
\begin{equation}
\psi_{2m+1}=(-1)^m\psi_1\prod_{k=1}^{m}\frac{J_{2k-1}}{J_{2k}}.
\label{eq:zero_mode_product}
\end{equation}
This proves uniqueness up to normalization and gives Eq.~\eqref{eq:log_zero_mode}.

To determine whether the logarithmic amplitude stays bounded as the chain grows, we examine the sum of adjacent log ratios in Eq.~\eqref{eq:log_zero_mode}. For \(0\leq\mu<t\), write \(q_p(\theta)=p\ln[t+\mu\cos\theta]\) and define

\begin{equation}
\Delta=\sqrt{t^2-\mu^2},
\qquad
s=\frac{\mu}{t+\Delta}.
\label{eq:Delta_r}
\end{equation}

These variables give the factorization
\begin{equation}
t+\mu\cos\theta
=
\frac{t+\Delta}{2}
\left(1+2s\cos\theta+s^2\right).
\label{eq:factorization}
\end{equation}

Taking the logarithm gives
\begin{equation}
q_p(\theta)
=
q_0
+
p\ln(1+2s\cos\theta+s^2),
\label{eq:qp_expanded}
\end{equation}

where
\begin{equation}
q_0
=
p\ln\frac{t+\Delta}{2}.
\label{eq:q0}
\end{equation}

To expand \(q_p\) in Fourier modes, we use the convergent series
\begin{equation}
\ln(1+se^{i\theta})
=
\sum_{k\ge1}
\frac{(-1)^{k+1}s^k}{k}e^{ik\theta},
\qquad |s|<1,
\label{eq:log_series}
\end{equation}

and substitute its real part into Eq.~\eqref{eq:qp_expanded} to obtain
\begin{equation}
q_p(\theta)
=
q_0+
2p\sum_{k\ge1}
\frac{(-1)^{k+1}s^k}{k}\cos(k\theta).
\label{eq:qp_fourier}
\end{equation}

We look for a coboundary function satisfying

\begin{equation}
\chi_p(\theta)-\chi_p(\theta-2\omega)
=
q_p(\theta-\omega)-q_p(\theta).
\label{eq:coboundary}
\end{equation}

The \(k\)-th Fourier mode of the right-hand side is

\begin{equation}
4p\frac{(-1)^{k+1}s^k}{k}
\sin\frac{k\omega}{2}
\sin\left[k\left(\theta-\frac{\omega}{2}\right)\right].
\label{eq:rhs_mode}
\end{equation}

Writing \(\chi_p\) in the Fourier form

\begin{equation}
\chi_p(\theta)
=
\sum_{k\ge1}
d_k
\cos\left[k\left(\theta+\frac{\omega}{2}\right)\right],
\label{eq:chi_fourier}
\end{equation}

we obtain

\begin{equation}
\chi_p(\theta)-\chi_p(\theta-2\omega)
=
-2\sum_{k\ge1}
d_k\sin(k\omega)
\sin\left[k\left(\theta-\frac{\omega}{2}\right)\right].
\label{eq:chi_difference}
\end{equation}

Equating coefficients of each Fourier mode gives

\begin{equation}
d_k
=
-p\frac{(-1)^{k+1}s^k}
{k\cos(k\omega/2)}.
\label{eq:ck}
\end{equation}

The zero-mean function is therefore
\begin{equation}
\chi_p(\theta)=
-p\sum_{k\ge1}
\frac{(-1)^{k+1}s^k}
{k\cos(k\omega/2)}
\cos\left[k\left(\theta+\frac{\omega}{2}\right)\right].
\label{eq:chi_p}
\end{equation}
For the golden-mean frequency, the reciprocal denominators grow at most polynomially with \(k\). Since \(s^k\) decays exponentially for \(\mu<t\), the series converges absolutely.

Summing Eq.~\eqref{eq:coboundary} at the phases \(\theta_{2k}\) gives \(S_m=\chi_p(\theta_{2m})-\chi_p(\theta_0)\). It follows that \(S_m\) is uniformly bounded. At \(\mu=0\), the hoppings are constant and \(S_m=0\).

For \(\mu\ge t\), \(\ln|J(\theta)|\) has logarithmic singularities at the zeros of \(J\). The set of these zeros is finite and nonempty, and it cannot be invariant under the irrational rotation by \(\omega=2\pi\alpha\). Thus \(\ln|J(\theta-\omega)|-\ln|J(\theta)|\) has at least one uncanceled singularity. The even-site phases \(\theta_{2m}\) are dense on the circle, so the increments \(S_m-S_{m-1}\) are unbounded in magnitude along a chain with no sampled zero bond. The range \(R(L)\) bounds the magnitude of each increment included in the chain and therefore grows without bound.

On the side \(\mu<t\), \(\chi_p\) also provides the transformation for the low-energy calculation. Equation~\eqref{eq:coboundary} implies that \(q_p(\theta)+\chi_p(\theta)+\chi_p(\theta-\omega)\), with \(q_p=\ln J\), is invariant under the irrational rotation by \(\omega\). This continuous function is therefore constant. Since \(\chi_p\) has zero mean, the constant is \(q_0=\langle q_p\rangle\). With \(g_n=e^{\chi_p(\theta_{n-1})}\), this gives \(J_ng_ng_{n+1}=e^{q_0}\). Multiplying the eigenvalue equation for \(\psi_n=g_n\eta_n\) by \(g_n\) gives Eq.~\eqref{eq:low_energy_gauge}.

The arithmetic--geometric mean inequality gives
\begin{equation}
\frac{e^{2\chi_p(\theta)}+e^{2\chi_p(\theta-\omega)}}{2}
\geq e^{\chi_p(\theta)+\chi_p(\theta-\omega)}
=\frac{e^{q_0}}{J(\theta)}.
\end{equation}
Taking the phase average yields \(C_p\geq\langle J^{-1}\rangle\). Near \(\theta=\pi\), the leading terms in \(t+\mu\cos\theta\) are \(t-\mu+\mu(\theta-\pi)^2/2\). Thus \(\langle J^{-1}\rangle\) diverges as \(\mu\to t^-\) for integer \(p\geq1\).

\section{Fourier coefficients and the large-\(p\) limit}

\label{sec:supp_fourier}

For the first-harmonic truncation in Sec.~\ref{sec:mu_FH}, the coefficients entering the self-duality condition are
\begin{equation}
\begin{aligned}
a_0(p,\mu)&=\frac{1}{2\pi}\int_0^{2\pi}
j(\theta)^p j(\theta+\omega)^p\,d\theta,\\
\widetilde V_1(p,\mu)&=\frac{1}{2\pi}\int_0^{2\pi}
\big[j(\theta)^{2p}+j(\theta-\omega)^{2p}\big]e^{-i\theta}\,d\theta.
\end{aligned}
\label{eq:a0_V1}
\end{equation}

To evaluate these integrals for integer powers, expand \(j(\theta)^q=\sum_{n=-q}^{q}c_n^{(q)}e^{in\theta}\). The identity
\begin{equation}
\cos^k\theta=2^{-k}\sum_{r=0}^{k}\binom{k}{r}e^{i(k-2r)\theta},
\label{eq:cos_power}
\end{equation}

gives the coefficients in the binomial expansion as
\begin{equation}
c_n^{(q)}=
\sum_{\substack{k=|n|,\ldots,q\\ k-|n|\ \mathrm{even}}}
\binom{q}{k}t^{q-k}\mu^k2^{-k}
\binom{k}{(k-|n|)/2},
\label{eq:cn}
\end{equation}
with \(c_{-n}^{(q)}=c_n^{(q)}\). The zeroth-order coefficient of \(a(\theta)\) is

\begin{equation}
a_0(p,\mu)=
\sum_{n=-p}^{p}c_n^{(p)}c_{-n}^{(p)}e^{-in\omega},
\label{eq:a0_sum}
\end{equation}

or, using \(c_{-n}^{(p)}=c_n^{(p)}\),
\begin{equation}
a_0(p,\mu)=
\left[c_0^{(p)}\right]^2
+
2\sum_{n=1}^{p}\left[c_n^{(p)}\right]^2\cos(n\omega).
\label{eq:a0_cos}
\end{equation}

For the effective diagonal term, the first-harmonic coefficient is

\begin{equation}
\widetilde V_1(p,\mu)=c_1^{(2p)}\left(1+e^{-i\omega}\right),
\label{eq:V1_p}
\end{equation}

and its magnitude is
\begin{equation}
|\widetilde V_1(p,\mu)|=
2\left|\cos\frac{\omega}{2}\right|\left|c_1^{(2p)}\right|.
\label{eq:V1_abs}
\end{equation}

For the quadratic modulation in Fig.~\ref{fig:ME_new}, the expansion is

\begin{equation}
j(\theta)^2=t^2+\frac{\mu^2}{2}+2t\mu\cos\theta+\frac{\mu^2}{2}\cos2\theta.
\label{eq:j2}
\end{equation}

Thus,

\begin{equation}
c_0^{(2)}=t^2+\frac{\mu^2}{2},
\qquad
c_1^{(2)}=t\mu,
\qquad
c_2^{(2)}=\frac{\mu^2}{4}.
\label{eq:c2}
\end{equation}

Substituting into \(a_0\) gives

\begin{equation}
a_0(2,\mu)=
\left(t^2+\frac{\mu^2}{2}\right)^2
+
2t^2\mu^2\cos\omega
+
\frac{\mu^4}{8}\cos2\omega.
\label{eq:a0_p2}
\end{equation}

On the other hand,

\begin{equation}
c_1^{(4)}=2t^3\mu+\frac32t\mu^3,
\label{eq:c1_4}
\end{equation}

so

\begin{equation}
|\widetilde V_1(2,\mu)|=
2\left|\cos\frac{\omega}{2}\right|
\left(2t^3\mu+\frac32t\mu^3\right).
\label{eq:V1_p2}
\end{equation}

Writing \(x=\mu/t\) and dividing both sides by \(t^4\), the self-duality condition becomes

\begin{equation}
\begin{aligned}
2\left|\cos\frac{\omega}{2}\right|
\left(2x+\frac32x^3\right)
&=\left(1+\frac{x^2}{2}\right)^2\\
&\quad +2x^2\cos\omega+\frac{x^4}{8}\cos2\omega.
\end{aligned}
\label{eq:crit_p2}
\end{equation}

For the golden mean, the positive root in the range considered is

\begin{equation}
\frac{\mu_c}{t}\approx0.514044.
\label{eq:mu_FH_p2}
\end{equation}

From \(j(\theta)=t[1+x\cos\theta]\) we obtain

\begin{equation}
a_0(p,\mu)=t^{2p}\mathcal A_p(x),
\qquad
|\widetilde V_1(p,\mu)|=t^{2p}\mathcal V_p(x).
\label{eq:scaling_t}
\end{equation}

The condition becomes

\begin{equation}
\mathcal V_p(x_c)=\mathcal A_p(x_c),
\label{eq:crit_scaled}
\end{equation}

where \(t^{2p}\) cancels completely. Therefore,

\begin{equation}
\frac{\mu_c}{t}=x_c(p,\alpha).
\label{eq:xc}
\end{equation}

Thus, for fixed \(\alpha\), the estimated ratio depends only on \(p\), and changing \(t\) only rescales the estimated modulation strength:

\begin{equation}
\mu_c(p,t)=t\,x_c(p).
\label{eq:mu_FH_scaling}
\end{equation}

For each integer \(p\), define

\begin{equation}
F_p(x)=|\widetilde V_1(p,tx)|-a_0(p,tx),
\label{eq:Fp}
\end{equation}

and solve \(F_p(x_c)=0\). For the golden mean, the resulting values are shown in Fig.~\ref{fig:powerlaw_mu_c}.

The truncation has no small parameter at these roots. For \(p=2,3,\) and \(4\), the root-mean-square spatial variation of \(a(\theta)\) is of the same order as its mean, and the second harmonic of \(V(\theta)\) is comparable to the first. Equation~\eqref{eq:crit_scaled} should therefore be interpreted as a characteristic first-harmonic scale whose accuracy must be assessed using the full-chain diagnostics.

The same coefficients also give the large-\(p\) behavior. To keep a finite modulation of \(J(\theta)/t^p\) as \(p\) increases, let

\begin{equation}
\frac{\mu}{t}=\frac{C}{p},
\label{eq:mu_C}
\end{equation}

with \(C\) finite. Then

\begin{equation}
j(\theta)^p=t^p\left(1+\frac{C}{p}\cos\theta\right)^p.
\label{eq:jp_C}
\end{equation}

As \(p\to\infty\),

\begin{equation}
\left(1+\frac{C}{p}\cos\theta\right)^p
\longrightarrow
e^{C\cos\theta}.
\label{eq:exp_limit}
\end{equation}

Therefore,

\begin{equation}
a(\theta)\sim t^{2p}e^{C\cos\theta+C\cos(\theta+\omega)}.
\label{eq:a_limit}
\end{equation}

Using

\begin{equation}
\cos\theta+\cos(\theta+\omega)=2\cos\frac{\omega}{2}\cos\left(\theta+\frac{\omega}{2}\right),
\label{eq:cos_sum}
\end{equation}

we obtain

\begin{equation}
\frac{a_0}{t^{2p}}\longrightarrow
I_0\left(2C\left|\cos\frac{\omega}{2}\right|\right),
\label{eq:a0_limit}
\end{equation}

where \(I_n\) is the modified Bessel function of the first kind. On the other hand,

\begin{equation}
V(\theta)\sim t^{2p}\left[e^{2C\cos\theta}+e^{2C\cos(\theta-\omega)}\right].
\label{eq:V_limit}
\end{equation}

Since

\begin{equation}
e^{z\cos\theta}=\sum_{n=-\infty}^{\infty}I_n(z)e^{in\theta},
\label{eq:bessel_expansion}
\end{equation}

the first-harmonic coefficient satisfies

\begin{equation}
\frac{|\widetilde V_1|}{t^{2p}}\longrightarrow
2\left|\cos\frac{\omega}{2}\right|I_1(2C).
\label{eq:V1_limit}
\end{equation}

The large-\(p\) limit of the self-duality condition is Eq.~\eqref{eq:large_p_closure}, with \(c_\alpha=|\cos(\omega/2)|=|\cos(\pi\alpha)|\). Expanding \(p\ln[1+(C/p)\cos\theta]\) shows that the error in Eq.~\eqref{eq:exp_limit} is \(O(p^{-1})\), uniformly in \(\theta\) for bounded \(C\). The two normalized Fourier coefficients have the same order of error. Since the limiting equation has a simple root at \(C_\infty\) for the golden-mean frequency, the scaled root \(p\mu_c/t\) differs from \(C_\infty\) by \(O(p^{-1})\), giving Eq.~\eqref{eq:mu_FH_asymptotic}.

\section{Numerical comparisons}

\label{sec:supp_comparison}

For the IPR comparison in Sec.~\ref{sec:p2_overview}, we use even periodic chains at \(p=2\) and \(t=1\) with Fibonacci approximants \(\alpha=377/610\) and \(1597/2584\). The reported IPRs are medians obtained by pooling the normalized eigenstates in each rank window at ring phases \(\phi=0.37\) and \(0.91\). For the Lyapunov calculation, we sample eigenenergies from the same windows at ring phase \(0.37\) and propagate transfer matrices at the irrational golden-mean frequency. The transfer variable \((\psi_n,J_{n-1}\psi_{n-1})^{\mathsf T}\) gives the step matrix
\begin{equation}
T_n(E)=
\begin{pmatrix}
E/J_n & -1/J_n\\
J_n & 0
\end{pmatrix}.
\end{equation}
We evaluate \(N^{-1}\ln\|T_N\cdots T_1\|_2\) per original site at \(N=8192,\ 32\,768,\ 131\,072\), average over the transfer phases \(0.13,\ 0.79,\ 1.78,\ 4.12\), and take the median over energies within each window. At the stronger of the two modulations compared in Sec.~\ref{sec:p2_overview}, the edge result is stable against both transfer length and the ring size used to select the energies, whereas the central result continues to decrease.

For the low-energy comparison, the largest level spacing in the lower half of the positive spectrum at \(\phi=0.37\) identifies the gap separating the central and outer groups. The central spectral fractions for the two Fibonacci sizes agree with the gap label \(\delta_0=\|2\alpha\|\). At moderate modulation, the width inferred from Eq.~\eqref{eq:Wc} is close to the numerical central half-width. Near the hopping-zero threshold it becomes much smaller than the full half-width, although a direct count in a narrow window around zero remains consistent with the local density estimate. This comparison separates the local DOS result from the additional arcsine model used to infer \(W_c\).

The spectra in Figs.~\ref{fig:ME_new} and~\ref{fig:phase_p34} use the even approximant \(L=610\) and \(\phi=0\). Checks with the next even Fibonacci size show the same change from inverse-size edge IPR to nearly size-independent edge IPR for \(p=2,3,\) and \(4\). The precise location of a finite-size FD crossing depends on the state window and should not be identified with an exact transition. The comparison with \(\mu_c\) is therefore made at the level of the modulation range and its trend with \(p\), rather than by matching several decimal-place crossing values.

\bibliography{references_1}

\end{document}